\documentclass[12pt]{article} 
\usepackage{graphicx}
\usepackage{textcomp} 
\usepackage{epstopdf}

\usepackage{latexsym}
\usepackage{amsmath}

\usepackage[mathcal]{euscript}
\usepackage{slashed}

\numberwithin{equation}{section}

\usepackage{marvosym}

\usepackage{indentfirst}

\newcommand\blfootnote[1]{
  \begingroup
  \renewcommand\thefootnote{}\footnote{#1}
  \addtocounter{footnote}{-1}
  \endgroup
}

\usepackage{hyperref}
\hypersetup{
        unicode,	
        colorlinks,
        citecolor=blue,
        linkcolor=blue, 
        urlcolor=red,
        bookmarksopen=true,
        bookmarksopenlevel=\maxdimen,
}

\usepackage{amsmath, amssymb}    

\numberwithin{equation}{section}

\usepackage{amsthm} 
\theoremstyle{definition}

\theoremstyle{plain}

\newtheorem*{result*}{Result}

\def\gl#1#2{\ifmmode \mathrm{GL}(#1; {\bf #2}) \else $\mathrm{GL}(#1; {\bf #2})$\fi}
\def\sl#1#2{\ifmmode \mathrm{SL}(#1; {\bf #2}) \else $\mathrm{SL}(#1; {\bf #2})$\fi}
\def\so#1{\ifmmode \mathrm{SO}({#1}) \else $\mathrm{SO}(#1)$\fi}

\def\sp#1#2{\ifmmode \mathrm{Sp}(#1; {\bf #2}) \else $\mathrm{Sp}(#1; {\bf #2})$\fi}
\def\usp#1{\ifmmode \mathrm{USp}(#1) \else $\mathrm{USp}(#1)$\fi}
\def\spin#1{\ifmmode \mathrm{Spin}(#1) \else $\mathrm{Spin}(#1)$\fi}
\def\su#1{\ifmmode \mathrm{SU}({#1}) \else $\mathrm{SU}(#1)$\fi}

\def\double #1{#1{\hbox{\kern-2pt $#1$}}}

\mathcode`\*="702A                  
\def\half{{\textstyle{1\over{\raise.1ex\hbox{$\scriptstyle{2}$}}}}}

\def \p{\partial}
\def \a{\alpha}
\def \b{\beta}
\def \e{\epsilon}
\def \d{\delta}

\def \l{\lambda}
\def \lb{\bar\lambda}
\def \L{\Lambda}
\def \hl{\hat\lambda}
\def \o{\omega}
\def \ob{\bar\omega}
\def \O{\Omega}
\def \t{\theta}

\def \ho{\hat\omega}

\def \hb{{\hat\b}}

\def \P{\nabla}
\def \St{{\rm STr}}
\def \st{{\text{\~Tr}}}
\def \tr{{\rm Tr}}
\def \Ib{{\mathbb I}}

\def \psu{{\mathfrak{psu}}}
\def \GL{\frac{GL(4|4)}{\left(GL(1)\times Sp(2) \right)^2}}

\begin{document}

\begin{flushright}
\makebox[0pt][b]{HU-EP-15/51}
\end{flushright}

\vspace{40pt}
\begin{center}
{\LARGE Supertwistor description of the $AdS$ pure spinor string}\\

\vspace{30pt}
{Israel Ram\'{\i}rez${}^{\clubsuit\diamondsuit}$ and
  Brenno Carlini Vallilo${}^{\spadesuit}$ }
\vspace{30pt}

{${}^{ \clubsuit}${\em Departamento de F\'isica, Universidad T\'ecnica Federico Santa Mar\'ia,\\
Casilla 110-V, Valpara\'iso, Chile}\\
\vskip 0.1in
 ${}^{ \diamondsuit}${\em Institut f\"ur Mathematik und Institut f\"ur Physik, Humboldt-Universit\"at zu Berlin\\
IRIS Haus, Zum Gro{\ss}en Windkanal 6,  12489 Berlin, Germany}\\
\vskip 0.1in
${}^{ \spadesuit}${\em Departamento de Ciencias F\'{\i}sicas,
Universidad Andres Bello, \\Republica 220, Santiago, Chile}}\\

\vspace{60pt}
{\bf Abstract}
\end{center}
We describe the pure spinor string in the $AdS_5\times S^5$ using
unconstrained matrices first used by Roiban and Siegel for the Green-Schwarz superstring.

\blfootnote{\\
${}^{\clubsuit}$
\href{mailto:ramirezkrause@gmail.com}{ramirezkrause@gmail.com}\\
${}^{\spadesuit}$ \href{mailto:vallilo@gmail.com}{vallilo@gmail.com}}

\setcounter{page}0
\thispagestyle{empty}

\newpage

\tableofcontents

\section{Introduction}

The superstring sigma model on $AdS$ spaces is usually described in terms 
of the supergroup coset $PSU(2,2|4)/SO(1,4)\times SO(5)$. The classical Green-Schwarz 
and pure spinor formulations are both well understood in terms of this coset. However for 
some applications, the usual exponential parametrization of the coset elements becomes cumbersome. 

In \cite{Roiban:2000yy} Roiban and Siegel introduced another parametrization for the 
$AdS_5\times S^5$ coset in terms of the supergroup $GL(4|4)$. The usefulness of this new formulation is in the fact that the coordinates can be represented in terms 
of unconstrained matrices. Furthermore, the coordinates transform in the fundamental representation of the superconformal group, like supertwistors. 

Depending on the application intended, different sets of coordinates are more useful than others. In the same way that global $AdS$ coordinates and Poincar\'e patch are useful for different applications. This also extends to the full superspace, {\em e.g.} chiral vs. nonchiral. The construction of explicit vertex operators for string states depends heavily on these choices. Since the beginning of the formalism vertex operators for $AdS$ have been discussed \cite{Berkovits:2000yr}. The first nontrivial example was introduced in \cite{Mikhailov:2009rx,Bedoya:2010qz}. Further developments can be found in \cite{Mikhailov:2011af, Chandia:2013kja}. The most complete description in the case of supergravity states was given in \cite{Berkovits:2012ps}. In this work the authors show that the ghost number two cohomology can be written in terms of harmonic superspace and a direct dictionary to the dual CFT single trace operators was obtained. The derivation is very lengthy due to the usual exponential parametrization of the coset elements. As advocated by Siegel \cite{Siegel:2010yd}, those results could be simplified using the $GL(4|4)$ description. This is one of the motivations to adapt the pure spinor formalism for this new coset. In this paper we will describe in detail how to achieve this. 

This paper is organized as follows. In Section 2 we describe the coset and its basic properties. In Section 3 the symmetries of $AdS$ are discussed in terms of the new coset. The full pure spinor superstring action is constructed in Section 4. In Section 5 we make a few comments on the construction of the vertex operator related to the $\beta$-deformations. In Section 6 we conclude the paper and discuss future lines of investigation. 

\section{The $GL(4|4) /(GL(1)\times Sp(2))^2$ coset}

Roiban and Siegel proposed a description of the $AdS_5\times S^5$
sigma model in terms of a coset that can be described by standard
matrices \cite{Roiban:2000yy}. The observation is that the
$PSU(2,2|4)$ group is a coset by itself (not caring about reality
conditions) $GL(4|4)/(GL(1)\times
GL(1))$, where the two $GL(1)$ groups are defined by scalar
multiplication in the upper and lower blocks. Note that the
super-determinant is invariant under the action of both $GL(1)$'s
combined.  Up to reality conditions ({\em i.e.} signature)
$AdS_5\times S^5$ can be described by \footnote{In our notation
  $Sp(n)$ describes $2n\times 2n$ matrices, {\em e.g.} $Sp(1)\simeq SU(2)$.}
\begin{equation}
 \frac{ GL(4|4)}{ (GL(1)\times Sp(2))^2}\,.
\end{equation}
Note that $Sp(2)=Spin(5)$ (under Wick rotation, $Sp(1,1)=Spin(1,4)$.)
Since we have a model with spinors, it is much more natural to work
with groups where the spinors transform in the fundamental representation.

The coset elements are denoted by $Z_M{}^A$ where the local
$\Lambda_A{}^B$ $(GL(1)\times Sp(2))^2$ transformations act on the
right by simple matrix multiplication. The index $M$ is a global
$GL(4|4)$ index. We divide both indices under bosonic and fermionic
elements $M=(m,\bar m)$ and $A=(a,\bar a)$. The $Sp(2)$ invariant
matrices will be denoted by $\Omega_{ab}$ and $\bar\Omega_{\bar a\bar
  b}$. There are analog matrices with indices up, which will be
denoted by the same symbol. They all satisfy $\Omega \Omega =
-{\mathbb I}$ where $\mathbb I$ is the identity matrix with
appropriate indices. We will omit explicit indices most of the time, only making them explicit when necessary.

The left-invariant currents (invariant under global transformations) are defined by
\begin{equation}
 J_A{}^B = Z_A{}^M d\, Z_M{}^B\,, \label{J}
\end{equation}
where $Z_A{}^M = (Z_M{}^A)^{-1}$.

A variation of the group element $Z$ around a background $Z_0$ is
given by
\begin{align}
 \delta Z_M{}^A= Z_M{}^B X_B{}^A\,,
\end{align}
where $X_B{}^A$ is given by
\begin{align}
 X_B{}^A = \left(\begin{array}{cc} X_b{}^a & \Theta_b{}^{\bar a}\\ \Theta'_{\bar b}{}^a & Y_{\bar b}{}^{\bar a}\end{array}\right). \label{X}
\end{align}

For these variations to be in the coset, the matrices $X$ and $Y$ must satisfy
\begin{equation}
 X^T = -\Omega\, X\,\Omega,\quad Y^T=- \bar\Omega\, Y\,\bar\Omega\,.
\end{equation}
Since these conditions do not imply that $X$ and $Y$ are traceless, we further impose
\begin{equation}
 {\rm Tr}\, X = {\rm Tr}\, Y=0\,.
\end{equation}
Doing this, we ensure that we work only with variations that are 
orthogonal to the gauge group.

We want to relate the elements described with the Roiban-Siegel formulation and the ones in the description using the $PSU(2,2|4)/(SO(5)\times SO(1,4))$ coset for the pure spinors. Our notation is closely related to the one adopted in \cite{Mazzucato:2011jt}. By construction, it is not hard to see the equivalence between $Z$ and the element $g\in PSU(2,2|4)/(SO(5)\times SO(1,4))$,
\begin{align}
 g\equiv Z\,.
\end{align}

In order to establish the equivalence between the content of the current in both formalism, we first identify the gauge content in our matrix formalism. Writing the block components of $J$ as
\begin{align}
 J = \left(\begin{array}{cc} J_X & K_1 \\ K_3 & J_Y\end{array}\right)\,, \label{J2}
\end{align}
we split the diagonal elements into three irreducible components using the $Sp(2)$ metric $\Omega$. Define for a matrix ${M_a}^{b}$ its three irreducible components,
\begin{align}
 \langle M \rangle =& \frac{1}{2}\left[M - \O M^T \O\right]-\frac{1}{4}{\mathbb I}{\rm Tr}\, M\,,\\
 \left( M \right) =& \frac{1}{2}\left[M + \O M^T \O\right]\,,\\
 {\rm Tr}\,M\,.
\end{align}
Usually, for
any matrix, one can split it in its antisymmetric, its symmetric traceless
and its trace part. Here $\langle M \rangle$ is the $\O$-antisymmetric,
$\O$-traceless part of $M$, $\left( M \right)$ is the $\O$-symmetric
part of $M$, and of course ${\rm Tr} M$ is the $\O$-trace of $M$. Using those independent structures, we can separate the element of $J$ \eqref{J} that are pure gauge.
we will define
\begin{align}
 K_X = \langle J_X\rangle\,, \quad A_X = (J_X)\,, \quad a_X= \frac{1}{4}{\rm Tr}J_X\,,\\
 K_Y = \langle J_Y\rangle\,, \quad A_Y = (J_Y)\,,\quad  a_Y=\frac{1}{4} {\rm Tr}J_Y\,.
\end{align}
$A_X$ and  $a_X$ are $Sp(2)$ and $GL(1)$ connections respectively. By definition,
\begin{align}
 J_X = K_X + A_X + {\mathbb I} a_x\,.
\end{align}
By checking its transformation property, we can now relate the diagonal elements in \eqref{J2} with the gauge part of current in the $\psu(2,2|4)$ algebra,
\begin{align}
 J_0^i\equiv \left(\begin{array}{cc} A_X + {\mathbb I} a_X & 0 \\ 0 & A_Y + {\mathbb I}a_Y\end{array}\right)\,.
\end{align}
The rest of the bosonic components are related as,
\begin{align}
 J_2^m \equiv \left(\begin{array}{cc} K_X & 0 \\ 0 & K_Y\end{array}\right)\,.
\end{align}

Before we continue, we have to make clear that the $\langle \cdot \rangle$ and $\left( \cdot \right)$ operations need to be treated with care when there is a product of fermionic matrices. Take two fermionic matrices $A$ and $B$, is easy to see that
\begin{align}
 {\rm Tr}\left(\frac{1}{2}\left[AB - \O B^T A^T \O\right]-\frac{1}{4}{\mathbb I}{\rm Tr} AB \right)=-{\rm Tr}AB\neq 0\,,\\
 {\rm Tr}\left(\frac{1}{2}\left[AB + \O B^T A^T \O\right]\right)={\rm Tr}AB\neq 0\,.
\end{align}
The solution to this problem is to add a $(-)$ sign when transposing fermionic matrices. Thus, for a product of two fermionic matrices $A$ and $B$, our three irreducible components read
\begin{align}
 \langle AB \rangle =& \frac{1}{2}\left[AB + \O B^T A^T\O\right]-\frac{1}{4}{\mathbb I}{\rm Tr} AB\,,\\
 \left( AB \right) =& \frac{1}{2}\left[AB - \O B^T A^T \O\right] \quad {\rm and} \quad {\rm Tr}AB\,.
\end{align}

It is not so obvious how to relate the fermionic part of $PSU(2,2|4)$, $J_1^\alpha$ and $J_3^{\hat\alpha}$, with the nondiagonal terms in \eqref{J2}, $K_1$ and $K_3$, because they do not have the right $\mathbb Z_4$ charge. The matrices which do have the right $\mathbb Z_4$ charge are $F_1$ and $F_3$ which define $K_1$ and $K_3$ as
\begin{align}
 K_1=\frac{1}{\sqrt 2}\left(F_1-F_3^*\right)E^{-1/4} \quad {\rm and} \quad K_3=\frac{1}{\sqrt 2}\left(F_1^*+F_3\right) E^{1/4}\,,
\end{align}
where
\begin{align}
 F_1^*= \bar \O F_1^T \O\,, \qquad F_3^*=\O F_3^T\bar \O\,,
\end{align}
and $E={\rm Sdet} Z$. Now the identification is
\begin{align}
 J_1^\a \equiv F_1\, \quad J_3^{\hat\alpha}\equiv F_3\,.
\end{align}

Following the same idea, we define the $\Theta$s as functions of elements with the right $\mathbb Z_4$ charge,
\begin{align}
 \Theta=\frac{1}{\sqrt 2}\left(\theta_1-\theta_3^*\right)E^{-1/4} \quad {\rm and} \quad \Theta'=\frac{1}{\sqrt 2}\left(\theta_1^*+\theta_3\right) E^{1/4}\,.
\end{align}

In the same way as we related the components of the currents generated by $g$ and $Z$, we can relate variations of $g$ and $Z$ by
\begin{align}
 x_2^m \equiv& \left(\begin{array}{cc} X & 0 \\ 0 & Y\end{array}\right)\,,\\
 x_1^\alpha \equiv& \theta_1\,,\\
 x_3^{\hat\alpha} \equiv& \theta_3\,.
\end{align}
It is easy to check that there are the correct number of bosonic and fermionic variations. 

Finally, we need to define the ghosts fields that are essential for the construction of the BRST operator. We define the right and left ghost, along with their conjugated momenta, as $\l_a^{~\bar a}$, $\o_{\bar a}^{~a}$, $\bar\l_{\bar a}^{~a}$, $\bar\o_a^{~\bar a}$. The indices in the ghost terms are such that $\l$ has the same indices as $F_1$ and $\bar\l$ the same as $F_3$. The crucial point to construct the right BRST operator is the pure spinor condition for $\l$ and $\bar\l$. Originally, the pure spinor condition was written in term of gamma matrices \cite{Berkovits:2000fe},
\begin{align}
\left(\l \gamma \l \right)^m=\left(\bar\l \gamma \bar\l \right)^m=0\,, \label{ps1}
\end{align}
which in turns implies
\begin{align}
 \l^\alpha \l^\b =&\frac{1}{16\cdot 5!}\gamma^{\alpha\b}_{mnopq}\left( \l \gamma^{mnopq}\l\right)\,,\qquad
 \bar\l^{\hat\alpha} \bar\l^\hb =&\frac{1}{16\cdot 5!}\gamma^{{\hat\alpha}\hb}_{mnopq}\left( \bar\l \gamma^{mnopq}\bar\l\right)\,. \label{psconstraint}
\end{align}
These constraints reduce the elements of $\l$ ($\lb$) from 16 to 11.

The pure spinor constraints in this matrix formulation read
\begin{align}
 \langle \l \l^* \rangle =0\,,&\qquad \langle \l^* \l \rangle=0\,,\label{psconstraint2}\\
 \langle \bar\l \bar\l^* \rangle=0\,,&\qquad \langle \bar\l^* \bar\l \rangle =0\,.
\end{align}
One can check that there are actually 5 constraints for $\l$($\lb$). Therefore,, our ghosts have 11 independent components, as expected. In a similar way to \eqref{psconstraint}, \eqref{psconstraint2} implies
\begin{align}
 {\l_a}^{\bar a} {\l_b}^{\bar b}=&-\frac{1}{16}\O_{ab}\bar\O^{\bar a\bar b}\tr\left[ \l\l^*\right]+{\l_{(a}}^{\bar a}{\l_{b)}}^{\bar b}+{\l_{\langle a }}^{\bar a}{\l_{b\rangle}}^{\bar b}\,,
\end{align}
and a similar condition for the $\bar\l$s. Note that
\begin{align}
 {\l_{(a}}^{\bar a}{\l_{b)}}^{\bar b}={\l_{a}}^{(\bar a}{\l_{b}}^{\bar b)}={\l_{(a}}^{(\bar a}{\l_{b)}}^{\bar b)}\,,
\end{align}
and the same is true for $\langle \rangle$.

The ghost Lorentz currents are defined as
\begin{align}
 N_X=&\frac{1}{2}\left(\l\o - \o^*\l^* \right)\,, \qquad \bar N_X=\frac{1}{2}\left(\bar\o\bar\l-\bar\l^*\bar\o^* \right)\,,\\
 N_Y=&\frac{1}{2}\left(\o\l - \l^*\o^* \right)\,, \qquad \bar N_Y=\frac{1}{2}\left(\bar\l\bar\o-\bar\o^*\bar\l^* \right)\,.
\end{align}
These definitions ensure that the $N$ and $\bar N$ terms transform as a gauge term.

Now we can make the identification between the ghost fields in the two descriptions:
\begin{align}
 \o_\alpha \equiv \o\,, \qquad \ho_{\hat\alpha} \equiv \bar\o\,, \qquad \l^\alpha\equiv \l\,,\qquad \hl\equiv \bar\l\,,\\
 N^i\equiv \left( \begin{array}{cc} N_X &0 \\ 0 & N_Y \end{array}\right) \quad {\rm and} \quad  \hat N^i\equiv \left( \begin{array}{cc} \bar N_X &0 \\ 0 & \bar N_Y \end{array}\right)\,.
\end{align}

\section{Symmetries of the $AdS$ Space}

The main aim of this article is to write a BRST-invariant superstring action embedded on a $AdS_5\times S^5$ target space in this formalism of unconstrained matrices. Since such action has to be invariant under the symmetries of a $AdS_5\times S^5$ space, we first proceed to understand how those symmetries act in this formalism and then we find the structures that are invariant under such symmetries.

\subsection{Local}

A local (gauge) transformation is given by
\begin{align}
 \delta_L Z= ZL+Z \left(\begin{array}{cc} \mathbb{I}\, \frac{l_X}{4} & 0 \\ 0 & \mathbb{I}\, \frac{l_Y}{4} \end{array}\right),
\end{align}
where
\begin{align}
 L=\left(\begin{array}{cc} L_X & 0 \\ 0 & L_Y\end{array}\right).
\end{align}
and $\left(L_{X/Y} \right)=L_{X/Y}$. The constraints for the $L$ matrices restrict them to be in $Sp(2)\times Sp(2)$, and the $l_X$ and $l_Y$ are the remaining terms of the stability group.

Thus, a local transformation on the current reads,
\begin{align}
 \delta_L \left( \begin{array}{cc} J_X & K_1\\ K_3 & J_Y \end{array}\right)= \left( \begin{array}{cc} \left[ J_X,L_X\right] + \p L_X + {\mathbb I}\frac{\p l_X}{4} & K_1 L_Y - L_X K_1 - K_1\frac{l_X-l_Y}{4}\\ K_3 L_X - L_Y K_3 +K_3\frac{l_X-l_Y}{4} & \left[ J_Y ,L_Y\right] + \p L_Y + {\mathbb I}\frac{\p l_Y}{4}   \end{array}\right).
\end{align}

Using that $L_X K_X= \O L_X^T K_X^T \O$ we find
\begin{align}
&\delta_L K_X = \left[ K_X,L_X\right]\,,& &\delta_L K_Y = \left[ K_Y,L_Y\right]\,, \\
&\delta_L A_X = \left[ A_X,L_X\right] + \p L_X\,, & &\delta_L A_Y = \left[ A_Y,L_Y\right] + \p L_Y\,, \\
&\delta_L a_X = \p l_X\,,&   &\delta_L a_Y = \p l_Y\,,\\
&\d_L F_1= - L_X F_1 + L_Y F_1\,, & &\d_L F_3=-L_Y F_3 - F_3 L_X\,,
\end{align}
which is expected due to the coset properties.

The first invariant structures that we find are
\begin{align}
 \d_L \tr \left[ K_X\bar K_X \right]= \d_L \tr \left[ K_Y\bar K_Y \right] = \d_L \tr \left[ K_1\bar K_3\right]=0\,.
\end{align}

The first attempt to construct a Wess-Zumino term will be to use $\left[K_1 \bar K_1^*\right]$ and $\left[K_3 \bar K_3^*\right]$. Note that the trace acts on two different spaces. It turns out that those structures are not invariants:
\begin{align}
 \d_L \ln{\rm Tr}\left[K_1 \bar K_1^*\right]=-\d_L \ln{\rm Tr}\left[K_3 \bar K_3^*\right]=-2\left(l_X-l_Y \right)\,.
\end{align}
To solve this issue we note that $\d_L E=\left(l_X-l_Y \right)E$. Therefore the right local invariant structures are
\begin{align}
 \d_L \tr \left[ K_1 \bar K_1^* E^{1/2}\right]=\d_L \tr \left[ K_3 \bar K_3^* E^{-1/2}\right]=0\,.
\end{align}

Since we are equipped with a gauge transformations we can define a covariant derivative,
\begin{align}
 \P Z=& \p Z - ZA -Za/4\,, \label{covariant}
\end{align}
where, as expected,
\begin{align}
 A=\left( \begin{array}{cc} A_X & 0 \\ 0& A_Y \end{array}\right), \qquad a=\left( \begin{array}{cc} \mathbb{I}a_X & 0 \\ 0& \mathbb{I}a_Y \end{array}\right),
\end{align}
and $\left(A_{X/Y} \right)=A_{X/Y}$.

Since $\left[ l,A\right]=\left[ l,a\right]=0$, is straightforward to show
\begin{align}
 \d_L \P Z=\P Z \left( L+l/4\right)\,.
\end{align}
This is the expected property for the covariant derivative. Finally, just to make everything explicit
\begin{align}
 \P Z^{-1}&= \p Z^{-1} + AZ + a Z/4\,,\\ \qquad \P E&=0\,.
\end{align}

The covariant derivative of the global invariant current is
\begin{align}\begin{split}
 \P J =& \p J -\left[ J,A+\frac{\Ib}{4}a\right]\\
      =&\left(\begin{array}{cc}\p J_X &\p K_1\\ \p K_3 & \p K_Y \end{array} \right)\\&+\left( \begin{array}{cc}\left[ A_X,J_X\right] & A_X K_1-K_1 A_Y +\frac{1}{4}\left(a_X-a_Y \right) K_1 \\  A_X K_3 -K_3 A_X -\frac{1}{4}\left(a_X-a_Y \right) K_3 & \left[ A_Y,J_Y\right]\end{array}\right).\end{split}
\end{align}
Thus, for the $F$s matrices we obtain
\begin{align}
 \P F_1 =& \p F_1 +A_X F_1 - F_1 A_Y\,,\\
 \P F_3 =& \p F_3 +A_Y F_3 - F_3 A_X\,.
\end{align}

For the ghosts we require that $\l$, $\bar\o$ behave as $F_1$, and $\bar \l$, $\o$ as $F_3$. The local invariance of $\tr\left[\o\bar \P \l\right]$ and $\tr\left[\bar\o \P \bar\l \right]$ requires that
\begin{align}
\d_L \l=& -L_X \l+\l L_Y\,,& \d_L \o =& - L_Y \o +\o L_X\,,\\
\d_L \bar \l=& -L_Y \bar\l+\bar\l L_X\,,& \d_L \bar\o =& - L_X \bar\o +\bar\o L_Y\,.
\end{align}

\subsection{Global}

As stated above, the currents $J$ are invariant under global transformations
\begin{align}
 \delta_G Z=MZ\,,
\end{align}
where $M$ is any global matrix. The ghosts fields are, by construction, invariant under global transformation, {\em i.e.}
\begin{align}
 \delta_G (\text{Ghosts})=0\,.
\end{align}

If we compute the global transformation under all the terms constructed above, we find that neither $\left[K_1 \bar K_1^* E^{1/2}\right]$ nor $\left[K_3 \bar K_3^* E^{-1/2}\right]$ are invariants for a general $M$:
\begin{align}
 \d_G \ln{\rm Tr}\left[K_1 \bar K_1^* E^{1/2}\right]=-\d_G \ln{\rm Tr}\left[K_3 \bar K_3^* E^{-1/2}\right]=\frac{1}{2}\St M\,.
\end{align}
Therefore, we require
\begin{align}
 {\rm STr} M=0\,.
\end{align}

\section{BRST transformation and BRST invariant action}

In \cite{Roiban:2000yy} the relation between the $GL$-formalism with the $PSU$-formalism constructed in \cite{Metsaev:1998it} of the Green-Schwarz superstring was established. So far we have established a relation between the elements of the pure spinor string \cite{Berkovits:2000fe} in both the $GL$-formalism and the $PSU$-formalism. We have also found all structures invariant under the global and local symmetries of the $AdS_5\times S^5$ space. In order to construct an action for the pure spinor superstring, we are missing one important ingredient the BRST operator. Below, we will establish the BRST symmetry and then find a BRST invariant action. Before doing so, we will review the BRST symmetry in the $PSU$-formalism. Then we will construct the BRST symmetry for the $GL$-formalism and construct the BRST invariant action, using the previous construction as a guide.

\subsection{$PSU$-formalism}

The BRST transformation for the group element is given by
\begin{align}
\e \d_B g=g\e \left(\l +\hl \right)\,. \label{brst0}
\end{align}
When acting on the global invariant current we obtain,
\begin{align}
 \e \d_B J=\p \e \left(\l+\hl \right) +\left[J, \e \left(\l +\hl \right) \right]\,.
\end{align}
It is useful to write the transformation for the different $\mathbb{Z}_4$-elements of the current,
\begin{align}
 \e \d_B J_0=&\left[J_1, \e \hl \right]+\left[J_3, \e \l \right]\,,\label{brst1}\\
 \e \d_B J_1=&\P \e \l +\left[J_2, \e \hl \right]\,,\\
 \e \d_B J_2=&\left[J_1, \e \l \right]+\left[J_3, \e \hl\right]\,,\\
 \e \d_B J_3=&\P \e \hl+\left[J_2, \e \l\right]\,,
\end{align}
where, as usual, the covariant derivative is defined as $\P=\p + \left[J_0, \right]$. The $\l$ and $\hat\l$ ghosts are invariants under the BRST transformation, but not the $\o$ and $\hat\o$. Thus the BRST transformation for the ghosts is given by,
\begin{align}
 \e\d_B \o =-J_3\e, & \qquad \e\d_B \l=0\,,\\
 \e\d_B \ho=-\bar J_1 \e, & \qquad\e\d_B \hl=0\,.\label{brst2}
\end{align}

The ghosts currents were already defined as\footnote{There is a minus sign of difference between our definition and the definition in \cite{Mazzucato:2011jt}.}
\begin{align}
 N=\left\lbrace \o,\l\right\rbrace \qquad {\rm and} \qquad \hat N=\left\lbrace \ho,\hl\right\rbrace\,.
\end{align}
Their BRST transformation are
\begin{align}
 \e\d_B N=-\left[J_3,\e\l \right] \qquad {\rm and} \qquad \e\d_B \hat N=-\left[\bar J_1,\e\hl \right]\,.
\end{align}

In order to prove the BRST invariance of the action we will use the Maurer-Cartan equations. They read
\begin{subequations}\label{MC}
\begin{align}
 \p \bar J_0 - \bar \p J_0 +\left[J_0, \bar J_0 \right]+ \left[J_1,\bar J_3 \right] + \left[J_2,\bar J_2 \right] + \left[J_3,\bar J_1\right]=0\,,\\
 \P \bar J_1 - \bar \P J_1 + \left[J_2,\bar J_3 \right] + \left[J_3,\bar J_2 \right]=0\,,\\
 \P \bar J_2 - \bar \P J_2 + \left[J_1,\bar J_1 \right] + \left[J_3,\bar J_3 \right]=0\,,\\
 \P \bar J_3 - \bar \P J_3 + \left[J_1,\bar J_2 \right] + \left[J_2,\bar J_1 \right]=0\,.
\end{align}\end{subequations}

Now we can show that the action
\begin{align}
 S_{\rm PSU}=\int d^2z \tr \left[\frac{1}{2}J_2\bar J_2 +\frac{1}{4}J_1 \bar J_3 + \frac{3}{4} \bar J_1 J_3 + \o \bar \P \l + \ho \P \hl -N\hat N \right]\,, \label{action1}
\end{align}
is BRST invariant.

Applying the BRST transformation given by \eqref{brst1}-\eqref{brst2} to \eqref{action1} we obtain,
\begin{align}\begin{split}
 \e\d_B S_{\rm PSU}=&\int d^2z \tr \left\lbrace\frac{1}{2}\left( \left[ J_1,\e\l\right] + \left[ J_3,\e\hl\right]\right) \bar J_2 + \frac{1}{2}\left( \left[\bar J_1,\e\l\right] + \left[ \bar J_3,\e\hl\right]\right) J_2 \right.\\
                    &+\frac{1}{4}\left(\P\e\l + \left[J_2,\e\hl \right]\right)\bar J_3 +\frac{1}{4}J_1 \left(\bar\P \e\hl + \left[ \bar J_2,\e\l\right] \right)  +\frac{3}{4}\left(\bar \P\e\l + \left[\bar J_2,\e\hl \right]\right) J_3 \\
                      &\left. + \frac{3}{4}\bar J_1 \left(\P \e\hl + \left[  J_2,\e\l\right] \right)-J_3\e \bar \P \l -\bar J_1 \e \P \hl +\o\left[\left[ \bar J_1, \e\hl\right]\right] \right.\\
                      & \left. + \left[\left[ \bar J_3,\e\l\right] ,\l\right] +\ho\left[\left[ J_1, \e\hl\right] + \left[ J_3,\e\l\right] ,\hl\right] + \left[ J_3,\e\l\right] \hat N + N \left[\bar J_1,\e\hl \right] \right\rbrace\\
                     =&\int d^2 z \tr \left\lbrace \frac{\e\l}{4}\left(\bar \P J_3- \P \bar J_3 + \left[\bar J_1,J_2 \right] - \left[ J_1,\bar J_2\right] \right)\right.\\
                      &\left. + \frac{\e\hl}{4}\left( \P \bar J_1 - \bar \P J_1 + \left[J_2,\bar J_3 \right] + \left[J_3,\bar J_2 \right] \right)-\e\l \left[ N,\bar J_3\right]-\e\hl \left[\hat N,J_1 \right] \right\rbrace\,.\end{split}
\end{align}
Using the pure spinor condition \eqref{ps1} and the Maurer-Cartan equations \eqref{MC} in the second equality, we can easily show,
\begin{align}
 \e\d_B S_{\rm usual}=0\,.
\end{align}

Before ending this section, we note that \eqref{brst0} is not actually nilpotent,
\begin{align}
 \e\d_B\e'\d_B g=g \e\e'(\l\l+\lb\lb+\lbrace\l,\lb\rbrace)\,.
\end{align}
Using the pure spinor condition \eqref{ps1} we can see that $ d_B^2\sim \lbrace \l,\lb\rbrace$. Therefore the BRST transformation is nilpotent up to a gauge transformation. The reason for this is that we are ignoring the BRST transformation for the ghosts. It was shown by Chand\'\i a in \cite{Chandia:2006ix} that in a general curved space the pure spinor ghosts acquire a nonvanishing BRST transformation. The case of $AdS$ background was discussed in more detail in \cite{Chandia:2014sta}. It is straightforward to adapt these results to the present case. 

\subsection{$GL$-formalism}

Now that we are familiar with the original BRST procedure, we can construct the right BRST transformation and the BRST invariant action using a $\GL$ coset. Our ansatz for the BRST transformation of $Z$ is
\begin{align}
 \e\d_B Z_M{}^A=Z_M{}^B \e\L_B{}^B \qquad \e\d_B Z_A{}^M=-\e \L_A{}^B Z_B{}^M\,. \label{brstgl1}
\end{align}
At first one would expect a $\L$ of the form
\begin{align}
 \L=\left(\begin{array}{cc}0&\l\\\lb &0\end{array} \right).
\end{align}
But a quick computation shows that $\d_B^2$ is not 0, nor even proportional to a gauge term. A correct form for $\L$ is
\begin{align}
 \L=&\left(\begin{array}{cc} 0 & \l_1\\ \l_3 & 0\end{array} \right)\,,\\
 \l_1=&\frac{1}{\sqrt 2}\left(\l-\bar\l^* \right)E^{-1/4}\,,\\
 \l_3=&\frac{1}{\sqrt 2}\left(\l^*+\bar\l \right)E^{1/4}\,.
\end{align}
This is of the right form since we want that $\e\d_B \ln E=\St \e\L=0$
, and also that $\d_B^2\sim gauge$, as discussed at the end of the previous subsection. Also, the form of $\l_1$ and $\l_3$ are such that $\l$ and $\lb$ transform as $F_1$ and $F_3$, respectively.

The transformation for the global invariant currents are
\begin{align}
 \e\d_B K_X=& \langle F_1\e\l^* + \e \bar \l^* F_3\rangle\,,\\
 \e\d_B A_X=& \left(F_1\e\bar\l-\e\l F_3 \right)\,,\\
 \e\d_B K_Y=& \langle F_1^*\e\l + \e \bar \l F_3^*\rangle\,,\\
 \e\d_B A_Y=& \left( F_3\e\l -\e\bar\l F_1\right)\,,\\
 \e\d_B F_1=&\P\e\l +\e\bar\l^* K_Y - K_X \e\bar \l ^*\,,\\
 \e\d_B F_3=&\P\e\bar\l - \e\l^* K_X + K_Y \e \l ^*\,,
\end{align}
and for the ghosts
\begin{align}
 \e\d_B \o =-\e F_3\,, \qquad \e\d_B \l=0\,,\\
 \e\d_B \bar\o =\e, \bar F_1\,, \qquad \e\d_B \bar \l=0\,.
\end{align}

Finally, we are able to show that
\begin{align}
 S_{\rm GL} = \int d^2 \st \left[\frac{1}{2} K_X \bar K_X - \frac{1}{2} K_Y \bar K_Y + \frac{1}{4}F_1 \bar F_3 + \frac{3}{4}\bar F_1 F_3 + \o \bar \P \l + \bar\o\P\bar\l +N_X\bar N_X- N_Y\bar N_Y   \right] \label{S}
\end{align}
is BRST invariant. Before we do that, a few comments are in order. $\st$ is defined in such a way to avoid confusion on which space the trace acts on. Since $\tr$ acts in either $a$ or $\bar a$ indices, we cannot write a term like $\tr\left(\l\o+\lb\ob\right)$. To avoid further confusion, we define an operation $\st$ such that $\st\left(\l\o+\lb\ob\right)$ means $ \l_a^{~\bar a}\o_{\bar a}^{~ a}+\lb_{\bar a}^{~a}\ob_{a}^{\bar a}$. Note that the trace of $K_Y$ has a minus sign, that is because $\St M=M_X-M_Y$. Also, while $\e\d_B\o$ has a minus sign, $\e\d_B \bar\o$ does not. That is because $F_3$ is related to $-J_3^{\hat\alpha}$, and we did that only for aesthetic reasons. Finally, in $S_{\rm usual}$ the ghost current term is $\tr-N\hat N$, and here is $\st N\bar N$. The difference in sign is because $\bar\o$ is equivalent to $-\ho_{\hat\alpha}$. In both actions we want that the kinetic term of the ghost to be positive defined. To obtain that, we need to define $\ho=-\ho_{\hat\alpha} \eta^{\alpha{\hat\alpha}}T_\alpha$, and this in turn implies that $\tr -N \hat N=N^i \hat N^j g_{ij}$ which is equivalent to $\St N\bar N$.

We are going to need the following Maurer-Cartan equation:
\begin{align}
 \bar \P F_1 - \P \bar F_1 - \bar K_X F_3^* + K_X \bar F_3^* -\bar F_3^* K_Y + F_3^* \bar K_Y=0,\\
 \bar \P F_3 - \P \bar F_3 + \bar K_Y F_1^* - K_Y \bar F_1^* +\bar F_1^* K_X - F_1^* \bar K_X=0.
\end{align}

We now check that

Applying the BRST transformation to \eqref{S},
\begin{align}\begin{split}
 \e\d_B &S_{\rm RS}=\int d^2z\st \left\lbrace \frac{1}{2}\left(F_1\e\l^* + \e\bar\l ^* F_3\right)\bar K_X +\frac{1}{2}K_X \left(\bar F_1\e\l^* + \e\bar\l ^*\bar F_3 \right) \right.\\
                   &- \frac{1}{2} \left( F_1^* \e\l + \e\bar \l F_3^*\right) \bar K_Y-\frac{1}{2} K_Y \left(\bar F_1^* \e\l + \e\bar \l \bar F_3^*\right) + \frac{1}{4}\left(\P \e\l + \e\bar\l^* K_Y - K_X \e\bar\l^* \right)\bar F_3 \\
                   &+ \frac{1}{4} F_1 \left( \bar\p\e\bar\l -\e\l^*\bar K_X +\bar K_Y \e\l^*\right)+\frac{3}{4}\left(\bar \P \e\l + \e\bar\l^*\bar K_Y -\bar K_X \e\bar\l^* \right) F_3 \\
                   &+ \frac{1}{4}\bar F_1 \left( \p\e\bar\l -\e\l^* K_X + K_Y \e\l^*\right)-\e F_3\bar\P\l + \e \bar F_1 \P \bar\l+\left(\bar F_1 \e\bar \l -\e\l \bar F_3 \right)N_X\\& -\left(F_1\e\bar\l - \e\l  F_3 \right)\bar N_X - \left(\bar F_3\e\l - \e\bar\l \bar F_1 \right) N_Y + \left(F_3\e\l - \e\bar\l F_1 \right)\bar N_Y\\
                   &\left. - \e\l F_3\bar N_X + N_X \e\bar F_1\bar\l + \e F_3 \l \bar N_Y - N_Y \bar\l \e \bar F_1 \right\rbrace.\end{split}
\end{align}
The pure spinor condition ensures that $ N_X\l - \l N_Y=0$ and $\bar N_Y\bar\l - \bar\l \bar N_X=0$. The only terms that survive are
\begin{align}
 \e\d_B S_{\rm RS}=&\int d^2z \frac{1}{4}\st \left[\e\bar\l\left( \bar \P F_1 - \P \bar F_1 - \bar K_X F_3^* + K_X \bar F_3^* -\bar F_3^* K_Y + F_3^* \bar K_Y \right) \right.\\
                   &\left. +\e\l\left(\bar \P F_3 - \P \bar F_3 + \bar K_Y F_1^* - K_Y \bar F_1^* +\bar F_1^* K_X - F_1^* \bar K_X \right) \right]\,,
\end{align}
which are identically 0 because of the Maurer-Cartan equation,
\begin{align}
 S_{WZ}=-\frac{1}{4}\int d^2z \tr \left[K_3^*\bar K_3 E^{-1/2}-K_1\bar K_1^*E^{1/2} \right]=-\frac{1}{4}\int d^2z \tr \left[ F_1\bar F_3 - \bar F_1 F_3\right]\,.
\end{align}

As we saw in the previous section, \eqref{S} is both local and global invariant if and only if the global transformation is generated by a supertraceless matrix.

\subsection{Vectors}

In \cite{Cagnazzo:2014yha,Candu:2013cga} a systematic construction of vertex operators for a supersphere sigma model was developed. An important ingredient for such construction was vectors describing the target spaced. The existence of such vectors describing the bosonic coordinates of the $Ads_5\times S^5$ superspace was discussed in \cite{Roiban:2000yy}. It is an interesting question whether we can construct all the matter part of \eqref{S} with such vectors. We will now discuss how to obtain this. The first set of vectors we can construct are
\begin{align}
 W^{MN}=Z_a{}^M\O^{ab}Z_b{}^N E^{1/4}\,, && W_{MN}= Z_M{}^a \O_{ab} Z_N{}^b E^{-1/4}\,,\label{v1}\\
 W'^{MN}=Z_{\bar a}{}^M\O^{\bar a\bar b}Z_{\bar b}{}^N E^{-1/4}\, , && W'_{MN}= Z_M{}^{\bar a} \O_{\bar a\bar b} Z_N{}^{\bar b} E^{1/4}\,.\label{v2}
\end{align}
Being careful with the indices and product of fermionic matrices, the only terms that we can construct are,
\begin{align}
 \P W^{MN} \bar\P W_{NM}=&\tr\left[ 4 K_X\bar K_X + 2 K_1\bar K_3 \right]\,,\label{v3}\\
 \P W'^{MN} \bar\P W'_{NM}=&\tr\left[ 4 K_Y\bar K_Y + 2 K_3\bar K_1 \right]\,.\label{v4}
\end{align}
Now we are able to construct part of the matter part of \eqref{S},
\begin{align}
 \frac{1}{8}\left[\P W^{MN} \bar\P W_{NM} -\P W'^{MN} \bar\P W'_{NM} \right]=& \st \left[\frac{1}{2} K_X\bar K_X - \frac{1}{2} K_Y\bar K_Y + \frac{1}{4} K_1\bar K_3 + \frac{1}{4} \bar K_1 K_3 \right]\,.
\end{align}

In order to obtain the right factor for the $K$-terms, we need to introduce another group of vectors,
\begin{align}
 U^{MN}=Z_a{}^M\O^{ab}\overleftrightarrow\P Z_b{}^N E^{1/4}\, , && U_{MN}= Z_M{}^a \O_{ab}\overleftrightarrow\P Z_N{}^b E^{-1/4}\, ,\\
 U'^{MN}=Z_{\bar a}{}^M\O^{\bar a\bar b}\overleftrightarrow\P Z_{\bar b}{}^N E^{-1/4}\, , && U'_{MN}= Z_M{}^{\bar a} \O_{\bar a\bar b}\overleftrightarrow\P Z_N{}^{\bar b} E^{1/4}\,.
\end{align}
We define $A\overleftrightarrow \P B=A\P B- \P A B $, and $\P$ is the covariant derivative defined in \eqref{covariant}. A direct computation shows that the product (in this case, the $\St$) of any two different vectors is always 0.

Using \eqref{v3} and \eqref{v4} we can construct,
\begin{align}
 U^{MN} \bar U_{NM}=&\tr\left[ -2 K_1\bar K_3\right]\,,\\
 \bar U'^{MN} U'_{MN}=&\tr\left[-2 K_3 \bar K_1 \right]\,.
\end{align}

With all those ingredients, we can construct the matter part of \eqref{S} without the Wess-Zumino term:
\begin{align}
 \frac{1}{8}\left[\P W^{MN} \bar\P W_{NM} - U^{MN}\bar U_{NM} -\P W'^{MN} \bar\P W'_{NM} + U'^{MN}\bar U'_{NM} \right]\\=\frac{1}{2}\st\left[ K_X\bar K_X - K_Y\bar K_Y + K_1\bar K_3 + \bar K_1 K_3 \right]\,.
\end{align}

The question now is how can we write the Wess-Zumino term of the action. First we remember that the Wess-Zumino term is
\begin{align}
 \mathcal{L}_{\rm WZ}=&-\frac{\kappa}{2}\tr\left[F_1 \bar F_3 - \bar F_1 F_3  \right]=\frac{\kappa}{2}\tr\left[K_1\bar K_1^* E^{1/2} - K_3^* \bar K_3 E^{-1/2} \right]\,.
\end{align}

A quick glance to list of vectors shows that the only possible way is a product between $W$s and $U$s. Indeed
\begin{align}
 (-)^M \P W^{MN}\bar U'_{NM}=&\tr\left[ -2K_1\bar K_1^* E^{1/2}\right]\,,\\
 (-)^M \P W'^{MN}\bar U_{NM}=&\tr\left[ -2K_3\bar K_3^* E^{-1/2}\right]\,.
\end{align}
Before we continue, a comment should be made: The product between vector is a STr between supermatrices,
\begin{align}
 W^{MN}W_{NM}=\St WW.
\end{align}
The product between $W$ and $U'$ should also be a STr. The product $\P W^{MN} \bar U'_{NM}$ is not, since
\begin{align}
 \P W^{MN} \bar U'_{NM}\neq \bar U'_{NM} \P W^{MN}\,.
\end{align}
The solution to this problem is the addition of the $(-)^M$ term. Now
\begin{align}
 (-)^M \P W^{MN} \bar U'_{NM} =\St \P W\bar U'\,.
\end{align}

We finally have all the ingredients to construct the matter part of $S_{\rm GL}$ and choosing $\kappa=\frac{1}{2}$, we get
\begin{align}
 \mathcal{L}_{\rm RS}=&\frac{1}{8}\left[ \P W^{MN} \bar\P W_{NM} - U^{MN}\bar U_{NM} - (-)^M \P W^{MN}\bar U'_{NM} -\P W'^{MN} \bar\P W'_{NM} \right.\nonumber\\
                      &\left. + U'^{MN}\bar U'_{NM} + (-)^M \P W'^{MN}\bar U_{NM} \right]\\
                     =&\frac{1}{2}\st\left[ K_X\bar K_X - K_Y\bar K_Y + \frac{1}{2}K_1\bar K_3 + \frac{3}{2}\bar K_1 K_3 \right]\,.
\end{align}

\section{An application: vertex operator construction}

Following \cite{Mikhailov:2009rx,Bedoya:2010qz} we will construct an operator $V$ such that $\e\d_B V=0$. To achieve this, we will construct the conserved current $j$ related to global symmetries of the action \eqref{S}. Then we will construct $V$ by applied the BRST transformation to $j$, $\d_B j=\partial V$. This will be our first vertex operator in this formalism. In future works we will try to apply the procedure explained in sections 4.3 to the construction of vertex operators, as in \cite{Cagnazzo:2014yha}.

\subsection{Equation of Motions}
As usual, in order to construct a conserved current, we need the equation of motions (EOM). To obtains such equations we will vary $Z$ around a background field,
\begin{align}
 Z=Z_0 e^X\,,
\end{align}
where the components of $X$ have been defined in \eqref{X}. This leads to
\begin{align}
 \d J= \p X + \left[ J,X\right]\, .
\end{align}
Writing this in components
\begin{subequations}
\begin{align}
 \d J_X=&\p X + \left[J_X,X \right] + K_1 \Theta_3 - \Theta_1 K_3\,,\\
 \d J_Y=&\p Y + \left[J_Y,Y \right] + K_3 \Theta_1 - \Theta_3 K_1\,,\\
 \d K_1=&\P \Theta_1 + K_X \Theta_1 -\Theta_1 K_Y + K_1 Y - X K_1\,,\\
 \d K_3=&\P \Theta_3 + K_Y \Theta_3 -\Theta_3 K_X + K_3 X - Y K_3\,.
\end{align}
\end{subequations}
Since we have written \eqref{S} in terms of $F_1$ and $F_3$, we write the variation of those, using the above equations:
\begin{align}
 \d F_1=& \P \theta_1 - K_X \theta_3^* + \theta_3^* K_Y - F_3^* Y + X F_3^*\,,\\
 \d F_3=& \P \t_3 + K_Y \t_1^* - \t_1^* K_X + F_1^* X - Y F_1^*\,,
\end{align}
and the same for the $K$s and $A$s:
\begin{align}
 \d K_X=& \P X + \frac{1}{2}\left( F_1\t_1^*-\t_1 F_1^* +\t_3^* F_3 - F_3^* \t_3\right)-\frac{\mathbb{I}}{4}\tr\left( F_1 \t_1^* + \t_3* F_3\right)\,,\\
 \d K_Y=& \P Y + \frac{1}{2}\left( F_1^*\t_1-\t_1^* F_1 +\t_3 F_3^* - F_3 \t_3^*\right)-\frac{\mathbb{I}}{4}\tr\left( F_1 \t_1^* + \t_3* F_3\right)\,,\\
 \d A_X=& \left[ K_X,X\right]+\frac{1}{2}\left( F_1\t_3+\t_3^* F_1^* - \t_1 F_3 - F_3^* \t_1^*\right)\,,\\
 \d A_Y=& \left[ K_Y,Y\right]+\frac{1}{2}\left( F_3\t_1+\t_1^* F_3^* - \t_3 F_1 - F_1^* \t_3^*\right)\,.
\end{align}

Using the variation of the action and the Maurer-Cartan equations we obtain,
\begin{subequations}\label{EOM}
\begin{align}
 \P \bar K_X + \frac{1}{2}\left(F_1\bar F_1^* - \bar F_1 F_1^* \right)- \frac{\mathbb I}{4} \tr\, F_1\bar F_1^* + \left[ \bar N_X, K_X\right]-\left[ N_X,\bar K_X \right]=&0 \,,\\
 \bar \P K_X + \frac{1}{2}\left(F_3^*\bar F_3 - \bar F_3^* F_3 \right)- \frac{\mathbb I}{4} \tr\, F_3\bar F_3^* + \left[ \bar N_X, K_X\right]-\left[ N_X,\bar K_X \right]=&0\,,\\
 \P \bar K_Y + \frac{1}{2}\left(F_1^*\bar F_1 - \bar F_1^* F_1 \right)- \frac{\mathbb I}{4} \tr\, F_1\bar F_1^* + \left[ \bar N_Y, K_Y\right]-\left[ N_Y,\bar K_Y \right]=&0\,,\\
 \bar \P K_Y + \frac{1}{2}\left(F_3\bar F_3^* - \bar F_3 F_3^* \right)- \frac{\mathbb I}{4} \tr\, F_3\bar F_3^* + \left[ \bar N_Y, K_Y\right]-\left[ N_Y,\bar K_Y \right]=&0\,,\\
 \bar \P \bar F_1 - \bar K_X F_3^* +K_X \bar F_3^* + F_3^* \bar K_Y - \bar F_3^* K_Y - N_X \bar F_1 + \bar N_X F_1 + \bar F_1 N_Y - F_1\bar N_Y=&0\,,\\
 \P F_1 - N_X \bar F_1 + \bar N_X F_1 + \bar F_1 N_Y - F_1\bar N_Y=&0\,,\\
 \P \bar F_3  -\bar K_Y^* F_1 + K_Y^* \bar F_1 + F_1^* \bar K_X - \bar F_1^* K_X - N_Y \bar F_3 + \bar N_Y F_3 + \bar N_Y F_3  - F_3\bar N_X=&0\,,\\
 \bar \P F_3 - N_Y \bar F_3 + \bar N_Y F_3 + \bar F_3 N_X - F_3\bar N_X=&0\,,\\
 \bar \P \o + \o \bar N_X - \bar N_Y \o=&0\,,\\
 \bar \P \l + \bar N_X \l - \l \bar N_Y=&0\,,\\
 \P \bar \o - N_X \bar \o + \bar \o N_Y=&0\,,\\
 \P \bar \l + \bar \l N_X - N_Y \bar \l=&0\,.
\end{align}
\end{subequations}
To obtain these equations we used the fact $\tr\left[ X H \right]=\tr\left[\langle X\rangle H \right]=\tr\left[X \langle H\rangle \right]$. Thus, the right EOM for $X$ is given by $\langle H\rangle=0$.

\subsection{Construction of $V$}

In order to compute the Noether current we first make a few observations. The first of them is noting that $\tr K_X \bar K_X=\tr K_X \bar J_X$, thus, instead of taking $\tr \bar K_X\langle Z_a{}^M\p M_M{}^N Z_N{}^b\rangle$, we just take $\tr \bar K_X Z_a{}^M\p M_M{}^N Z_N{}^b$. The same can be done for the ghost current, since $N_X=\left( \l\o \right)$. Using the EOM \eqref{EOM} and the global transformation studied in section 3.2, we find the left and right conserved currents,
\begin{align}
 j=&\left( \begin{array}{cc} K_X+2N_X &  \frac{1}{2\sqrt 2}\left(F_1-3F_3^* \right)E^{-1/4} \\ \frac{1}{2\sqrt 2}\left(F_1^*+3F_3^* \right)E^{1/4} & K_Y+2N_Y \end{array}\right)\,,\\
 \bar j=&\left( \begin{array}{cc} \bar K_X - 2\bar N_X &  \frac{1}{2\sqrt 2}\left(3\bar F_1-\bar F_3^* \right)E^{-1/4} \\ \frac{1}{2\sqrt 2}\left(3\bar F_1^*+\bar F_3^* \right)E^{1/4} & \bar K_Y- 2\bar N_Y \end{array}\right)\,.
\end{align}

 Since $\e\d_B \d_G S_{\rm RS}=0$ one would expect $\e\d_B j=\p V$ and $\e\d_B \bar j=-\bar \p V$ as in the usual description. But here $\St M=0$, thus, $\e\d_B j=\p V+\mathbb{I} A$ and $\e\d_B \bar j=-\bar \p V+ \mathbb{I} B$ is the most general form, for any $A$ and $B$. For the same reason, one would expect that $\e\d_B \e'\d_B \d_G S=0$ yields $\e\d_B V=0$, but the most general possibility is $\e\d_B V=\mathbb{I} C$, for any $C$. Now, this $\mathbb{I}C$ term should be expected from the gauge group $\left( GL(1)\right)^2$, since a the condition $A=\O A^T \O$, imposed to gauge terms, does not apply to the term proportional to the trace\footnote{Note that $\mathbb I=-\O \mathbb{I}^T\O$.} thus, it seems that we have eliminated those term. But this is not true, we did eliminated the $a_X$, $a_Y$ gauge terms: we did it when writing the action proportional to the $\tr$. Therefore, the correct BRST invariant vector is $\St V$.

After a long calculation, for BRST transformation of the left current we find that
\begin{align}
 \e\d_B j=& \frac{1}{2\sqrt 2}\p \e V-\frac{\mathbb I}{4}\tr\left(F_1 \e\l^*+\e\bar\l^* F_3 \right)\,,\\
 \e V=&Z\left(\begin{array}{cc} 0 & \e\left(\l+\bar\l^* \right)E^{-1/4}\\ \e\left(\l^* - \bar\l \right)E^{1/4} & 0 \end{array} \right)Z^{-1}=Z\e\L' Z^{-1}\,.
\end{align}
For the right current we find, as expected,
\begin{align}
 \e\d_B \bar j=& -\frac{1}{2\sqrt 2}\bar \p \e V-\frac{\mathbb I}{4}\tr\left(\bar F_1 \e\l^*+\e\bar\l^*\bar  F_3 \right)\,.
\end{align}

Finally, we check that $\e\d_B \St V=0$:
\begin{align}
 \e'\d_B \e V=&Z\left[ \e' \L,\e\L'\right]Z^{-1}=Z\e'\e \left\{ \L,\L'\right\}Z^{-1}\\
             =&2\e'\e Z\left(\begin{array}{cc} \l\l^* + \bar\l^* \bar\l & 0 \\ 0 & \l^*\l +\bar \l \bar\l^*\end{array}\right)Z^{-1}\\
             =&\frac{1}{2}\e'\e \mathbb{I}\tr\left(\l\l^* + \bar\l^* \bar\l \right),
\end{align}
therefore $\d_B \St V=0$. The vertex operator corresponding to the $\beta$-deformation discussed in \cite{Mikhailov:2009rx,Bedoya:2010qz} can now be described as the tensor product of two $V$. 


\section{Conclusion and further directions}

We have described the pure spinor superstring in $AdS_5\times S^5$ using the $GL(4|4)/(Sp(2)\times GL(1))^2$ coset first used by Roiban and Siegel for the Green-Schwarz superstring in 
\cite{Roiban:2000yy}. This formulation provides additional choices for the parametrization of the $AdS$ coordinates. This additional choices have been shown to be useful in formulations 
different superspaces relevant to the $AdS/CFT$ conjecture \cite{Siegel:2010yd}. Recently, Schwarz described another parametrization for the GS string in  $AdS_5\times S^5$ \cite{Schwarz:2015lla}. As was shown by Siegel \cite{Siegel:2015qka}, this new formulation can also be used in the present case. 

Furthermore, the complete superspace propagator for the entire tower of Kaluza-Klein modes was calculated in \cite{Dai:2009zg} using this new coset. This propagator was shown to be 
invariant under $\kappa$-symmetry. Since there is a close relation between $\kappa$-symmetry and the BRST transformations of the pure spinor formalism \footnote{For example, demanding invariance under $\kappa$-symmetry of the GS action in a general curved supergravity background puts the background on-shell. The same is achieved in pure spinor formalism demanding BRST invariance \cite{Berkovits:2001ue}.} it is likely that this propagator can be used to construct a BRST invariant ghost number two superspace function. Such function would be related to the unintegrated vertex operators of the supergravity modes in the pure spinor formulation. We are presently working in this direction. The ultimate goal is to have a systematic way to construct vertex operators at any mass level using the world sheet dilatation operator \cite{Ramirez:2015rma} to derive physical state conditions. Although BRST invariance should also be imposed, vanishing world sheet anomalous dimension may be enough to calculate the spacetime energy of the string states.

\section*{Acknowledgments}

We would like to thank Osvaldo Chand\'\i a and
William Linch for useful discussion. We are especially grateful to
Luca Mazzucato for collaboration during initial stages of this work.
The work of I.R. is supported by CONICYT project No. 21120105 and the USM-
DGIP PIIC grant. The work of B.C.V. is partially supported by FONDECYT grant number
1151409 and CONICYT grant number DPI20140115.

\newpage

\bibliography{mybib}{}
\bibliographystyle{abe}

\end{document}